\ifx\undefined\documentclass
\documentstyle[a4,epsf]{elsart}
\newcommand\mathcal[1]{{\cal #1}}
\newcommand\lesssim{\stackrel{\lower.7ex\hbox{$<$}}{\lower.7ex\hbox{$\sim$}}}
\else
\documentclass[a4paper]{elsart}
\usepackage{epsf}
\usepackage{amssymb}
\fi
\makeatletter
\ifx\undefined\operator@font
  \let\operator@font=\rm
\fi
\def\Re{\mathop{\operator@font Re}\nolimits}
\def\Im{\mathop{\operator@font Im}\nolimits}
\makeatother

\begin{document}
\thispagestyle{empty}


\begin{flushright}
INLO-PUB-01/96\\
NIKHEF 96/04
\end{flushright}

\vspace{2cm}

\begin{frontmatter}
\title{Definitions and upper bounds\\for unstable boson masses}

\author{Wim~Beenakker${}^1$, Geert Jan van Oldenborgh${}^2$}
\address{Instituut-Lorentz, Rijksuniversiteit Leiden, Netherlands}
\author{Jiri~Hoogland${}^2$, Ronald~Kleiss${}^2$}
\address{NIKHEF-H, Amsterdam, Netherlands}
\thanks{research made possible by a fellowship of the Royal Dutch Academy of 
Arts and Sciences}
\thanks{research supported by the Stichting FOM}
\begin{abstract}
We consider the problem of a proper definition of the mass of a heavy unstable 
boson.  It is shown how various definitions are related by the 
renormalization-scheme independence of the full resummed amplitude.  This is 
made explicit for the $W$ boson in the large $N_f$ limit, where we find an 
upper bound on the peak position of the lineshape and on the mass, provided a 
sensible definition of the mass is used.  In more realistic models, we 
conjecture that under certain conditions the mass is similarly bounded.  Using 
this approach an upper limit of about 1 TeV on the mass of the 
Minimal-Standard-Model Higgs boson is found, independent of any breakdown of 
perturbation theory.
\end{abstract}
\journal{Phys.\ Lett.\ B}
\date{January 1996}
\end{frontmatter}
\clearpage


\section{Introduction}

The problem of a proper definition of the mass of unstable bosons has been 
hotly debated \cite{Zphysics,Stuart1,ScottZ} in the past.  For the $W$ and $Z$ 
bosons, the differences between the various definitions are perturbative and 
easily calculable, as $\Gamma/m \approx 1/40$.  It is thus a matter of 
convention which definition is used.  The scheme preferred at LEP1 is to 
renormalize the mass at real values of $p^2=s$, and we shall call this the 
LEP1-scheme.  The alternative scheme, which we shall dub the complex-pole (cp) 
scheme, uses the position of the pole in the complex $s$ plane as the 
renormalization point.  The LEP1 scheme gives rise to a running width in the 
propagator, whereas the cp scheme yields a fixed width.

The difference between the various schemes becomes especially relevant 
in the case of a heavy Higgs boson, which will be looked for by the next 
generation of proton colliders \cite{LHC}.  Its width can be a sizable 
fraction of the mass: $\Gamma_H/m_H \approx
m_H^2/(1425\mbox{ GeV})^2$ perturbatively.   The behaviour $\Gamma \propto 
m^3$ is general to a theory with massive bosons and a fixed low-energy 
effective coupling.

In this letter we investigate the various definitions of the mass of an 
unstable boson, and how these are bounded by the $m^3$-behaviour of the width. 
We perform this investigation in a model, where we can resum the perturbation 
theory exactly: a $W$ boson with $G_\mu\to0, N_d\to\infty$ ($N_d$ is the 
number of fermion doublets) such that $G_\mu N_d$ is finite.  We also discuss 
whether the occurrence of a Landau pole jeopardizes these results.  Finally, 
we conjecture a generalization of the mass bound and apply this to
the Standard-Model $W$ and Higgs boson.


\section{Results for $W$ bosons in the large $N_d$ limit}
\label{sec:largeNd}

To investigate the behaviour of an unstable boson with a very large width, we 
study the reaction $e^+\nu_e\to\mu^+\nu_\mu$ in a model with $N_d\to\infty$ 
massless fermion doublets and a coupling constant $G_\mu\to0$, such that their 
product stays finite.  In this renormalon approximation, all one-loop 
irreducible corrections are given by the one-loop fermionic self-energy 
diagram.  All other diagrams are suppressed by powers of $1/N_d$.  Up to 
spinorial factors, the unrenormalized amplitude is then given by
\begin{equation}
    \mathcal{A}(s) = {\hat{g}^2 s \over s - \hat{\mu} + \hat{g}^2\Sigma(s)}
\;,
\end{equation}
where the real quantities $\hat{g}$ and $\hat{\mu}$ indicate the 
unrenormalized coupling constant and mass squared, respectively.  The one-loop 
fermionic contribution to the $W$ self-energy is given by
\begin{equation}
\label{eq:defSigma}
    \Sigma(s) = \frac{N_d}{6\pi^2}s\left(\Lambda - \log(s+i\epsilon) + 
        i\pi\right)
\qquad (\epsilon\downarrow0)
\;.
\end{equation}
As is well-known, the non-zero imaginary part implies a finite life-time for 
the $W$ boson.  The infinity of the self-energy is parameterized by the real 
quantity $\Lambda$, and all physical results are independent of it.

We now introduce a class of mass-renormalization schemes by the following 
condition on the scaled inverse amplitude $V(s) = s/\mathcal{A}(s)$:
\begin{equation}
\label{eq:massrenorm}
    \Re V(\mu) = 0
    \quad\Rightarrow\quad
    \hat{\mu} = \Re\mu + \hat{g}^2\Re\Sigma(\mu)
\end{equation}
for some (complex) renormalization point $\mu\equiv m^2(1-ix)$.  The LEP1 
scheme is given by the specialization $\Im\mu = 0$ and $\mu = m_r^2$, the cp 
scheme by $\Im V(\mu) = 0$ and $\mu=\mu_c=m_c^2(1-ix_c)$.  Intermediate 
schemes can be defined to connect these two special cases.

An essential ingredient in our discussion is the existence of an effective 
low-energy coupling constant $G_\mu$, embodied in the Fermi condition 
\begin{equation}
    \frac{\sqrt{2}}{G_\mu} \equiv - V(0) = \frac{\hat\mu}{\hat{g}^2} = 
        \frac{\Re\mu}{\hat{g}^2} + \Re\Sigma(\mu)
\;.
\end{equation}
Note that $G_\mu$ is real.  We thus have
\begin{equation}
\label{eq:defV}
    V(s) = -\frac{\sqrt{2}}{G_\mu} + \Sigma(s) + \frac{s}{\Re\mu}
        \left( \frac{\sqrt{2}}{G_\mu} - \Re\Sigma(\mu) \right)
\;,
\end{equation}
which is independent of $\Lambda$.  The natural scale in this problem is
\begin{equation}
    A = {6\pi\sqrt{2} \over N_dG_\mu} \;.
\end{equation}
Perturbatively this yields a decay width $\Gamma = m^3/A$.
In the Minimal Standard Model we find $A = (437\mbox{ GeV})^2$ for $N_d=12$ 
({\em i.e.\/}, neglecting all fermion masses).   We may now write
\begin{eqnarray}
\label{eq:Vs}
    V(s) & = & - \frac{N_d}{6\pi}\left( A + 
        \frac{s}{\pi}\log\frac{s+i\epsilon}{A} - is - s\Phi(\mu) \right)
\;,\\
    \hspace*{-3ex}\Phi\bigl(m^2(1-ix)\bigr) & = & \frac{A}{m^2} + 
        \frac{1}{\pi}\log\frac{m^2}{A} 
        + \frac{1}{2\pi}\log(1+x^2) 
        - x\bigl(1 + \frac{\arctan x}{\pi} \bigr)
\;.
\end{eqnarray}
Note that we have traded $\hat{g}$ and $\hat{\mu}$ for $A$ and $\Phi$.  Lines 
in the complex $\mu$ plane with $\Phi(\mu)$ constant correspond to exactly the 
same physics\footnote{It is easy to verify that the amplitudes corresponding 
to $V(s)$ are unitary.}.  The LEP1 renormalization scheme corresponds to the 
intersect of these lines with the real axis, $x=0$; the cp scheme, 
$V(\mu_c)=0$, implies
\begin{equation}
\label{eq:complexpole}
    m_c^2 = \frac{Ax_c}{(1+x_c^2)(1+\arctan(x_c)/\pi)}
\;.
\end{equation}

\begin{figure}[tbp]
\begin{center}
\unitlength 1pt
\begin{picture}(390,272)(0,0)
\put(0,0){\epsffile{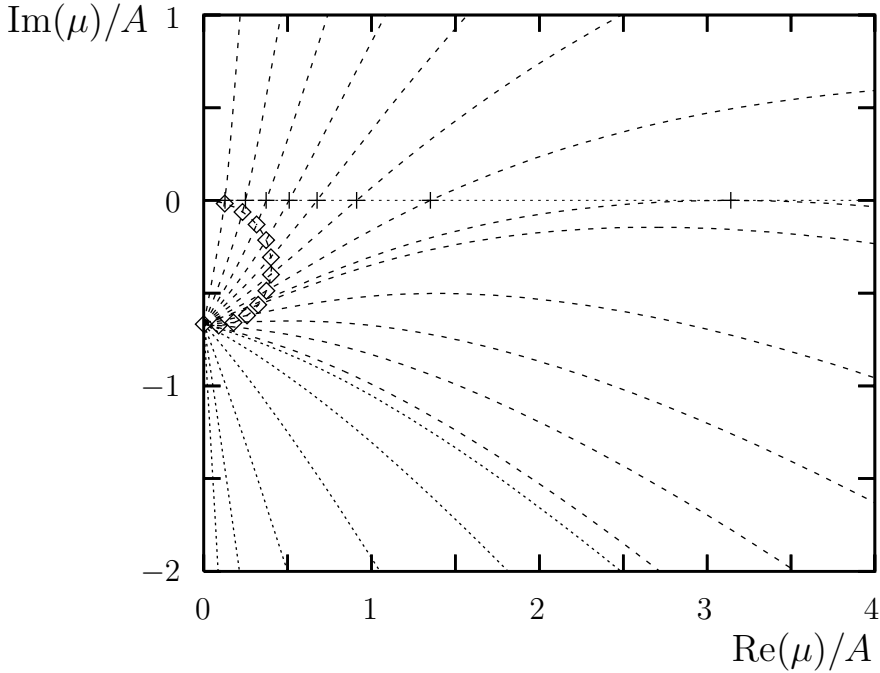}}
\put(337,5){\makebox(0,0)[r]{\large$\Re(\mu)/A$}}
\put(10,250){\makebox(0,0)[tl]{\large$\Im(\mu)/A$}}
\put(75.6, 35.1){\makebox(0,0)[r]{$-2$}}
\put(75.6,105.0){\makebox(0,0)[r]{$-1$}}
\put(75.6,175.0){\makebox(0,0)[r]{$0$}}
\put(75.6,244.9){\makebox(0,0)[r]{$1$}}
\put( 84.0,21.1){\makebox(0,0){$0$}}
\put(147.2,21.1){\makebox(0,0){$1$}}
\put(210.5,21.1){\makebox(0,0){$2$}}
\put(273.7,21.1){\makebox(0,0){$3$}}
\put(336.9,21.1){\makebox(0,0){$4$}}
\end{picture}
\end{center}
\caption[]{Lines of equal physics in the $\mu$ plane.  The intersection with 
the real axis defines the LEP1 renormalization scheme, the intersect with the 
egg-shaped curve at the left the cp scheme.}
\label{fig:mucurves}
\end{figure}

In Fig.~\ref{fig:mucurves} we present some lines of constant $\Phi$ for 
$\Re\mu>0$.  One sees that for $\Phi < (1+\log(\pi))/\pi \approx 0.6827$ there 
are no longer any real solutions, so one cannot describe a $W$ boson with this 
mass in the LEP1 scheme.  This limiting value is given by $m_r^2=\pi A$, or 
equivalently $x_c=\Gamma_c/m_c \approx 1.605$ and $m_c^2=0.3393\,A$.  For 
larger $\Phi$, there are two solutions for $m_r^2$, of which the small one is 
the usual one.  It is amusing to note that within this renormalon 
approximation a value $m_W = 1.392\;10^{22}$ GeV would give precisely the same 
lineshape as the standard $W$ mass of $80.23$ GeV.  In the region with dotted 
lines with $\Phi < \log(2/3)/\pi \approx 0.2122$ ($\mu_c/A=-2i/3$) the 
propagator does not retain a pole, and the cp scheme is ill-defined. However, 
it still describes a bona-fide lineshape.

At this point we want to remark the following.  It could be argued that a 
large value of $x_c = \Gamma_c/m_c$ indicates a break-down of either 
perturbation theory or the particle interpretation of the propagating field.  
However, we see no signs of such a breakdown, even for $x_c\to\infty$. It 
remains to be seen whether the complete disappearance of a singularity in the 
propagator {\em does\/} indicate a limitation of the perturbative approach.  
The disappearance of the pole can be interpreted as the change from a 
propagating, albeit very short-lived, particle to a non-propagating particle 
that mediates an effective interaction.

The mass of the $W$ is thus bounded in both renormalization schemes.  In the 
LEP1 scheme this is an unphysical effect --- some perfectly acceptable values 
for $\Phi$ give lines that simply do not intersect the real axis.  In 
contrast, the absolute value of $\mu_c$ is properly defined on a much larger 
parameter space. Within this parameter space it peaks at $x_c \approx 4.493$ 
($|\mu_c| \approx 0.6825\,A$).

It remains to study how these bounds are related to the peak position in the 
experimentally measurable spectrum, which is the physically relevant mass 
definition.  The lineshape is proportional to
\begin{equation}
    \sigma(s) = \left(\frac{N_d}{6\pi}\right)^2 \frac{s}{|V(s)|^2}
\;.
\end{equation}

\begin{figure}[tbp]
\begin{center}
\unitlength 1pt
\begin{picture}(390,272)(0,0)
\put(0,0){\epsffile{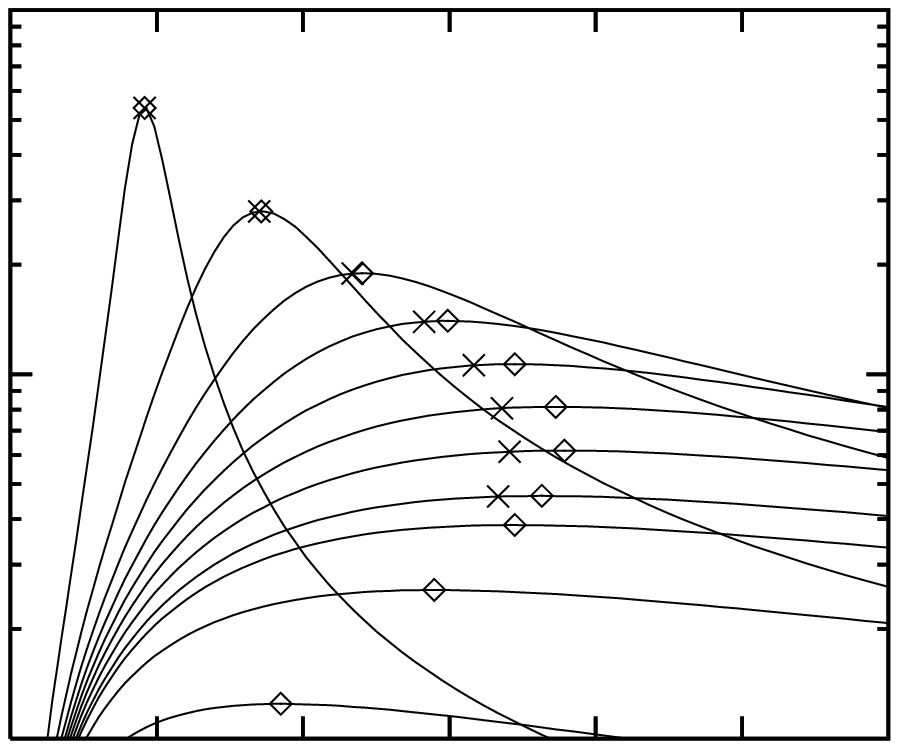}}
\put(337,0){\makebox(0,0)[r]{\large$s/A$}}
\put(30,250){\makebox(0,0)[tl]{\large$\sigma A$}}
\put(75.6, 35.1){\makebox(0,0)[r]{$0.1$}}
\put(75.6,140.0){\makebox(0,0)[r]{$1$}}
\put(75.6,244.9){\makebox(0,0)[r]{$10$}}
\put( 84.0,21.1){\makebox(0,0){$0.0$}}
\put(168.3,21.1){\makebox(0,0){$0.4$}}
\put(252.6,21.1){\makebox(0,0){$0.8$}}
\put(336.9,21.1){\makebox(0,0){$1.2$}}
\end{picture}
\end{center}
\caption{The lineshape for various values of $x_c$.  The positions of the 
peaks are indicated by diamonds, the values of $|\mu_c|$ by crosses.}
\label{fig:ampli}
\end{figure}

In Fig.\ \ref{fig:ampli} we show this quantity for some values of $x_c$.  
Somewhat surprisingly, we find that the peak position is also bounded, and has 
qualitatively the same behaviour as $|\mu_c|$.  Indeed, in an {\em effective\/} 
model with an inverse amplitude $V(s) = (s-\mu_c)/g^2$ (with constant $g$) one 
would find $s_{\mathrm{peak}} = |\mu_c|$.  Upon inclusion of all higher-order 
effects, the peak position $s_{\mathrm{peak}}$ is given implicitly by
\begin{equation}
    \Phi = \Psi^\pm(s_{\mathrm{peak}})
\;,\quad
    \Psi^\pm(s) = \frac{1}{\pi}\Bigl(1 + \log\frac{s}{A}\Bigr)
        \pm \sqrt{\Bigl(\frac{\pi A - s}{\pi s}\Bigr)^2 - 1}
\;,
\end{equation}
where one chooses $\Psi^\pm$ when $\Phi \stackrel{\scriptstyle>}{\scriptstyle<} 
\Psi^+(s_{\mathrm{peak}}^{\mathrm{max}}) = 
\Psi^-(s_{\mathrm{peak}}^{\mathrm{max}})$, with
\begin{equation}
\label{eq:smaxmax}
    s_{\mathrm{peak}}^{\mathrm{max}} = \frac{\pi}{\pi + 1}A
\;.
\end{equation}
This last equation gives the largest attainable peak position, 
$s_{\mathrm{peak}}^{\mathrm{max}} \approx 0.7585\,A$ for $\mu_c = 0.6824\,A$ 
and $x_c=4.165$.  For the Standard Model $W$ boson, 
$s_{\mathrm{peak}}^{\mathrm{max}} \approx (380\;\mathrm{GeV})^2$.

\begin{figure}
\begin{center}
\unitlength 1pt
\begin{picture}(390,272)(0,0)
\put(0,0){\epsffile{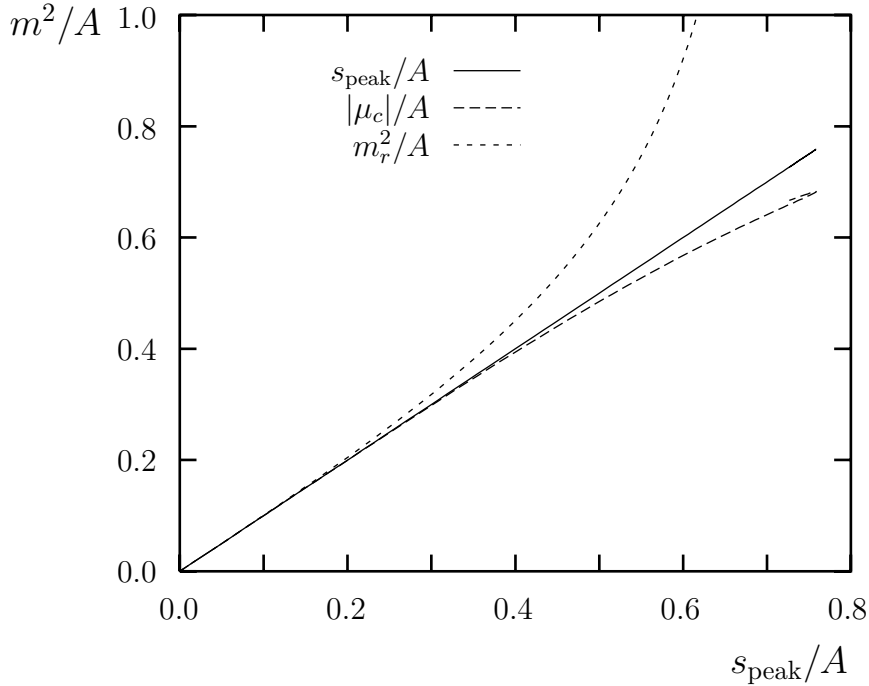}}
\put(337,0){\makebox(0,0)[r]{\large$s_{\mathrm{peak}}/A$}}
\put(20,250){\makebox(0,0)[tl]{\large$m^2/A$}}
\put(178.8,223.9){\makebox(0,0)[r]{$s_{\mathrm{peak}}/A$}}
\put(178.8,209.9){\makebox(0,0)[r]{$|\mu_c|/A$}}
\put(178.8,195.9){\makebox(0,0)[r]{$m^2_r/A$}}
\put(75.6, 35.1){\makebox(0,0)[r]{$0.0$}}
\put(75.6, 77.1){\makebox(0,0)[r]{$0.2$}}
\put(75.6,119.0){\makebox(0,0)[r]{$0.4$}}
\put(75.6,161.0){\makebox(0,0)[r]{$0.6$}}
\put(75.6,202.9){\makebox(0,0)[r]{$0.8$}}
\put(75.6,244.9){\makebox(0,0)[r]{$1.0$}}
\put( 84.0,21.1){\makebox(0,0){$0.0$}}
\put(147.2,21.1){\makebox(0,0){$0.2$}}
\put(210.5,21.1){\makebox(0,0){$0.4$}}
\put(273.7,21.1){\makebox(0,0){$0.6$}}
\put(336.9,21.1){\makebox(0,0){$0.8$}}
\end{picture}
\end{center}
\caption{Relation of various mass definitions to the peak position in the 
lineshape.}
\label{fig:masses}
\end{figure}

We are now in a position to compare the different definitions of the mass of 
an unstable boson.  In Fig.\ \ref{fig:masses} we show the absolute value of 
the complex mass and the LEP1 mass as a function of the peak position.  As we 
suspected the absolute value of the complex mass gives the best indication of 
the peak position, up to very high widths.  The LEP1 renormalization scheme 
breaks down very quickly, and also deviates substantially below this.  We note 
that the maximum value reached by the absolute value of the complex mass 
($0.6825\,A$) is fairly close to the maximum of the peak position ($0.7585\,
A$).  However, for $\Phi<\log(2/3)/\pi$ the complex mass is also ill-defined 
and the only definition left is the peak position, if we accept these 
solutions as legitimate.


\section{Landau poles}

\begin{figure}[tbp]
\begin{center}
\unitlength 1pt
\begin{picture}(390,272)(-30,-20)
{
\renewcommand\epsfsize[2]{0.5#1}\put(0,0){\epsffile{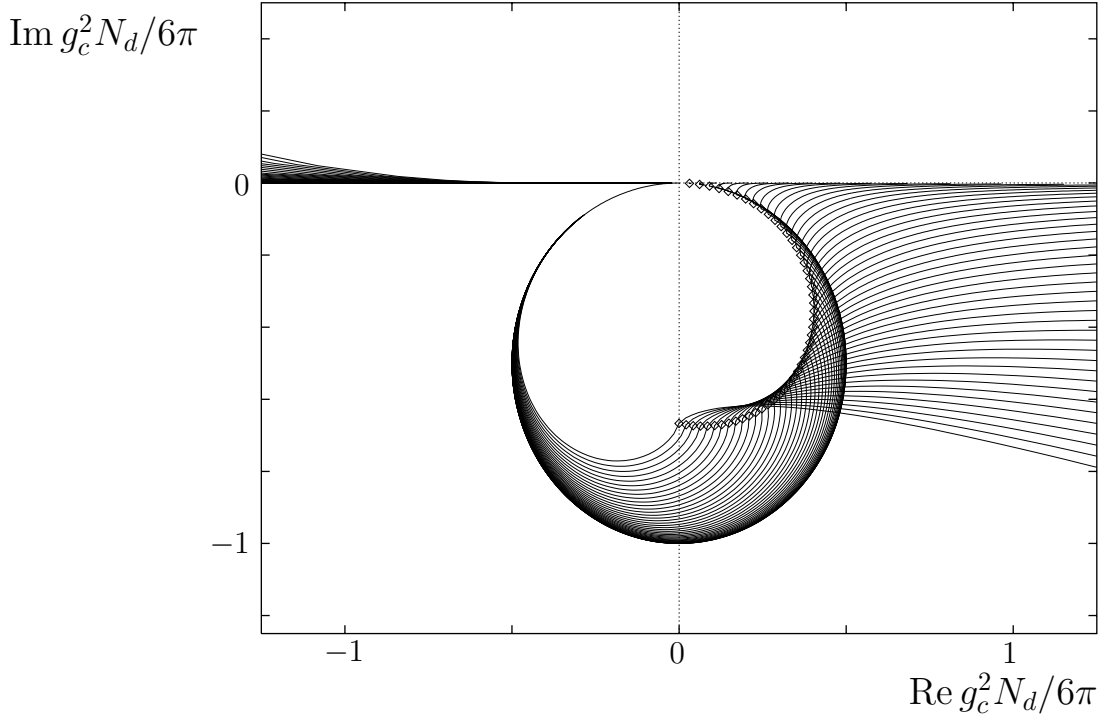}}
\put(350,-12){\makebox(0,0)[r]{\large$\Re g_c^2 N_d/6\pi$}}
\put(10,245){\makebox(0,0)[tr]{\large$\Im g_c^2 N_d/6\pi$}}
\unitlength 0.5pt
\put(58.8, 89.1){\makebox(0,0)[r]{$-1$}}
\put(58.8,361.0){\makebox(0,0)[r]{$0$}}
\put(130.2,7.1){\makebox(0,0){$-1$}}
\put(382.1,7.1){\makebox(0,0){$0$}}
\put(633.9,7.1){\makebox(0,0){$1$}}
}
\end{picture}
\end{center}
\caption[]{The running of the coupling in the cp scheme.  
The sickle in the center is generated for $s>0$, the diamonds for $s=0$, and 
the fans starting out to the right and returning from the left for $s<0$.}
\label{fig:g2s}
\end{figure}

In the previous section, we have always assumed that the theory is valid 
in the peak region.  However, the occurrence of a Landau pole at the same
scale would indicate that the renormalon approximation we are using is no 
longer adequate.  Using the cp scheme, we can write the inverse amplitude as 
\begin{equation}
    V(s) = \frac{s-\mu_c}{g_c^2(s)}
\;,\quad 
    \frac{1}{g_c^2(s)} = \frac{1}{\hat{g}^2} + 
            \frac{\Sigma(s)-\Sigma(\mu_c)}{s-\mu_c}
\;.
\end{equation}
For positive $s$, the running coupling is given by
\begin{equation}
\label{eq:defg2s}
    \frac{1}{g_c^2(s)} = \frac{1}{g_c^2(0)}\left( 1 - \frac{s\mu_c}{\pi A} 
                \frac{\log(s/\mu_c)}{s-\mu_c} \right)
\;.
\end{equation}
The Landau pole would be situated at the point where $g_c^2(s)$ diverges:
\begin{equation}
    A\pi\left(\frac{1}{\mu_c} - \frac{1}{s}\right) = \log\frac{s}{\mu_c}
\;.
\end{equation}
Equating the imaginary parts and using Eq.~(\ref{eq:complexpole}), 
we find that there is no solution for real $s>0$.
The Landau pole for positive $s$ has been moved into the complex plane.
This can also be seen from Fig.~\ref{fig:g2s};
the running coupling is bounded by 
$|N_d g_c^2(s)/6\pi + i/2| < 1/2$ for positive $s$.
In the LEP1 renormalization scheme $g_r^2(s)$ {\em does\/} exhibit a pole for 
$s>0$, but the amplitude stays finite, as required by the scheme-independence.

\begin{figure}[tbp]
\begin{center}
\unitlength 1pt
\begin{picture}(390,272)(0,0)
\put(0,0){\epsffile{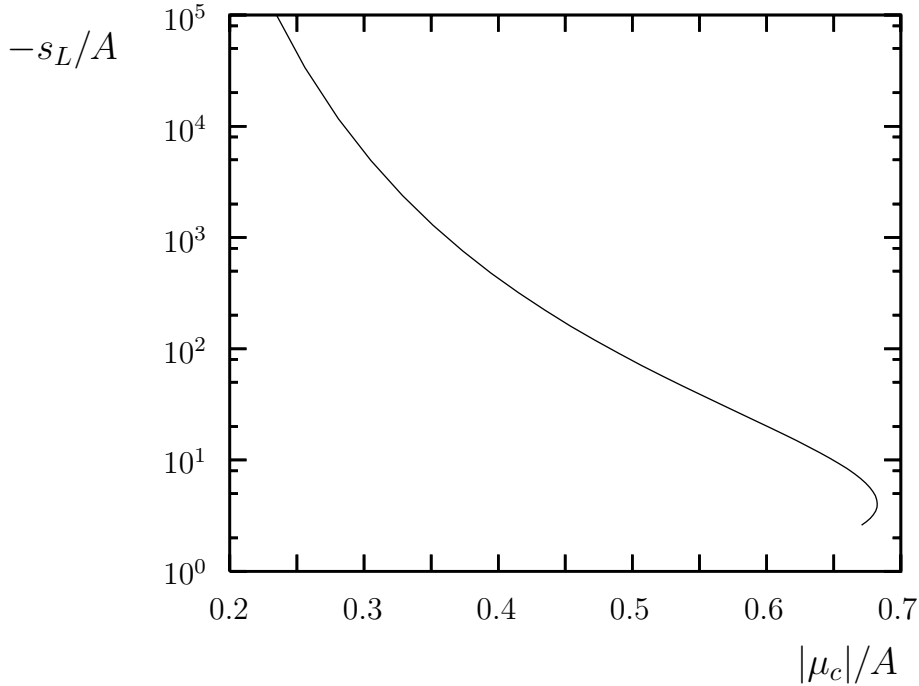}}
\put(337,0){\makebox(0,0)[r]{\large$|\mu_c|/A$}}
\put(0,240){\makebox(0,0)[tl]{\large$-s_L/A$}}
\put(75.6, 35.1){\makebox(0,0)[r]{$10^0$}}
\put(75.6, 77.1){\makebox(0,0)[r]{$10^1$}}
\put(75.6,119.0){\makebox(0,0)[r]{$10^2$}}
\put(75.6,161.0){\makebox(0,0)[r]{$10^3$}}
\put(75.6,202.9){\makebox(0,0)[r]{$10^4$}}
\put(75.6,244.9){\makebox(0,0)[r]{$10^5$}}
\put( 84.0,21.1){\makebox(0,0){$0.2$}}
\put(134.6,21.1){\makebox(0,0){$0.3$}}
\put(185.2,21.1){\makebox(0,0){$0.4$}}
\put(235.7,21.1){\makebox(0,0){$0.5$}}
\put(286.3,21.1){\makebox(0,0){$0.6$}}
\put(336.9,21.1){\makebox(0,0){$0.7$}}
\end{picture}
\end{center}
\caption[]{The position of the Landau pole for $s<0$ as a function of the 
approximate peak position $|\mu_c|$.}
\label{fig:landau}
\end{figure}

For negative $s$, one has to be careful about the continuation of the 
logarithms in Eq.~(\ref{eq:defg2s}), which becomes
\begin{equation}
\label{eq:defg2sm}
    \frac{1}{g_c^2(s)} = \frac{1}{g_c^2(0)}\left( 1 - \frac{s\mu_c}{\pi A} 
        \frac{\log(-s/\mu_c)+i\pi}{s-\mu_c} \right)
\;.
\end{equation}
As required by the optical theorem, the amplitude now is real; 
this follows trivially from Eq.~(\ref{eq:Vs}).
In this region we {\em do\/} encounter a Landau singularity; both $g_c^2(s_L)$ 
diverges, and the inverse amplitude $V(s_L)$ vanishes, for some $s_L<0$.  In 
Fig.\ \ref{fig:landau} we show the position of the Landau pole as a function 
of the mass.  One can see that the singularity is always a relatively safe 
distance away from the mass scale: at the largest mass, $|\mu_c|=0.6825\,A$, 
we have $s_L \approx -5.9\,|\mu_c|$, while for $x_c\to\infty$, $s_L \approx 
-4.4\,|\mu_c|$.  We therefore conclude that effects from the Landau pole are 
not likely to influence the conclusion that the mass is bounded.

\section{Beyond the renormalon}

Up to this point we have considered only the renormalon approximation 
$N_d\to\infty$.  In the Standard Model the fermion loops form only part of the 
description of massive bosons.  Bosonic contributions to the self-energy, 
though not directly contributing to the on-shell width, may alter the picture 
considerably.  Unfortunately, the inclusion of these is not straightforward.  
We do not know all  higher-order corrections, and therefore problems arise 
when one resums the propagator, but computes the other corrections to a fixed 
order: a seperation between these two sets is in general not gauge invariant.

Some features of the resummed amplitude can however be inferred without 
explicit calculations.  We conjecture that the mass $|\mu_c|$ is bounded under 
certain conditions, and give an estimate for its upper value.

The conditions are
\begin{itemize}
\item The theory contains a heavy unstable boson $B$, which decays into light 
particles  ($m_i \ll m_B$).
\item All other particles have masses which are of the same scale as the $B$ 
mass.
\item There is a low-energy condition, resulting in a dimensionful effective 
coupling constant, so that the perturbative width is proportional to $m_B^3$.
\end{itemize}

We can take over the analysis of Eqs (\ref{eq:massrenorm}-\ref{eq:defV}), 
except that in general $\Sigma(0)\neq0$.  A non-zero imaginary part of 
$\Sigma(0)$ forces 
the introduction of a phase in the Fermi condition, $\sqrt{2}/G_\mu\,\equiv 
-V(0) \exp(i\phi) $.  This phase is not due to the existence of open decay 
channels, but to the use of a complex renormalized mass inside loops.  
We obtain
\begin{eqnarray}
V(s) & = & -\frac{\sqrt{2}}{G_\mu}\cos\phi + \Sigma(s) - \Re\Sigma(0)
\nonumber\\&&\mbox{}
            - \frac{s}{\Re\mu}\Bigl( -\frac{\sqrt{2}}{G_\mu}\cos\phi + 
                \Re\Sigma(\mu) - \Re\Sigma(0) \Bigr)
\;,
\end{eqnarray}
with
\begin{equation}
    \sin\phi = \Im\Sigma(0) \frac{G_\mu}{\sqrt{2}}
\;.
\end{equation}
The position of the complex pole $\mu_c$ is given by $\Im V(\mu_c)=0$.  As a 
function of $x_c = -\Im\mu_c/\Re\mu_c$ we find
\begin{equation}
\label{eq:mc2}
    m_c^2(1+x_c^2)\Im\left( \Sigma(\mu_c)\over\mu_c \right) = x_c 
        \left( {\sqrt{2}\over G_\mu} \cos\phi + \Re\Sigma(0) \right)
\;.
\end{equation}

Under the conditions listed above, the most important features of the one-loop 
self-energy $\Sigma(s)$ are
\begin{eqnarray}
\label{eq:selfform}
    \Sigma(s) & = & - \frac{\beta_0}{16\pi^2} \Bigl(s\Lambda + i \pi sf - 
        s \log s + \alpha_1 s \log\frac{\mu_c}{s} + \alpha_2 \Re F(\mu_c)
\nonumber\\&&\phantom{-\frac{\beta_0}{16\pi^2} \Bigl(}
        + \mbox{neglected terms} \Bigr) \hspace{15ex} (\Re s>0)
\;,
\end{eqnarray}
where $\beta_0$ is the first coefficient of the $\beta$-function which governs 
the evolution of the coupling parameter $g^2(s)$.  
In our convention, the $\beta$-function is given by
\begin{equation}
    16\pi^2 \frac{d g^2}{d\log Q^2_{{}_\mathrm{UV}}} = -\beta_0\, g^4 
        + \mathcal{O}(g^6)
\;.
\end{equation}
By $f$ we denote the fraction of the $\beta$-function that corresponds to the 
light particles into which the bosons can decay.  In the model considered 
before, $\beta_0 = -8N_d/3$ and $f=1$.  The arbitrary real numbers $\alpha_i$ 
do not influence the results.  The neglected terms contain various small 
logarithms, plus a term $\Lambda\Im\mu$ which is formally infinite but must 
disappear when higher orders are taken into account.

Combining Eqs (\ref{eq:mc2}) and \ (\ref{eq:selfform}), we find
\begin{equation}
    m_c^2 = - \frac{16\pi}{\beta_0} \frac{\sqrt{2}}{G_\mu} 
        \frac{x_c}{(1+x_c^2)[f + \arctan(x_c)/\pi]}
\;.
\end{equation}
This gives egg-shaped curves with a maximum in $|\mu_c|$ near $x_c\to\infty$, 
\begin{equation}
    \lim_{x_c\to\infty} |\mu_c| = - \frac{16\pi}{\beta_0} 
        \frac{\sqrt{2}}{G_\mu} \frac{1}{f + 1/2}
\;.
\end{equation}


\section{Applications to various models}

Assuming the approach of the previous section (in particular the form of the 
self-energy) to be justified, we now apply it to a few models of unstable 
bosons. 

In the first place, we can apply this to a naive, effective description of 
an unstable boson, in which $\Sigma(s) = is\gamma\theta(s)$.  
This is the description used, for instance, in the 
`improved Born approximation' to the $Z^0$ lineshape at LEP1 energies.
It gives rise to a complex pole on the semicircle 
\begin{equation}
        \Bigl|\mu_c \frac{G_\mu}{\sqrt{2}} \gamma + {i\over2}\Bigr| = {1\over2}
\;.
\end{equation}

Secondly, for the Standard-Model $W$ boson, we have \cite{Vaughn&Machacek} 
$\beta_0 = 8 \, (43/6 - N_d/3)$, $f\beta_0 = -8N_d/3$.  The mass is thus 
limited by $|\mu_c| \lesssim 4N_dA/(6N_d - 43) \approx (560\mbox{ GeV})^2$, 
again using $N_d=12$.  Preliminary studies of the explicit inclusion of the 
one-loop bosonic corrections give a very similar result.  Note that the Landau 
pole has disappeared in this asymptotically free theory.  Moreover, in the 
limit $N_d\to\infty$, we recover the results of section \ref{sec:largeNd}.

Finally, we turn to the Standard-Model Higgs boson, only taking into account 
the bosonic corrections.   
The dimensionless coupling constant is $\lambda$, which we define through 
its low-energy behaviour in the reaction $W_L^+ W_L^- \to Z_L Z_L$, with 
$m_{W,Z}^2 \ll s,|\mu_H|$:
\begin{equation}
    \lambda = {\mu_H \over v^2}
\quad \Rightarrow \quad 
    V(0) = v^2
\;.
\end{equation}
Note that $v^2$ now plays the role of $\sqrt{2}/G_\mu$ in the $W$-boson 
analysis.  This is possible because of the low-energy theorem, which 
guarantees that the amplitude is given by $\mathcal{A}(s) = s/v^2$ for 
$s\ll|\mu_H|$ \cite{Chanowitz&Gaillard}.  We use a (tadpole) renormalization 
scheme in which $v = 246$ GeV is an observable and is not subject to 
higher-order corrections within our approximations \cite{Veltman&Yndurain}.

Due to gauge cancellations the self-energy has the form (\ref{eq:selfform}) 
with $\beta_0 = -3$ \cite{Vaughn&Machacek}.  From the standard perturbative 
expression for the width of the Higgs boson we infer $A_H = m_H^3/\Gamma_H = 
32\pi v^2/3$, this gives $f=1/2$.  Therefore, the large-width value of 
$|\mu_H|^{1/2}$ is conjectured to be 1000 GeV, and the upper bound on the 
Higgs mass and peak position should be close to this value.


\section{Summary and conclusions}

In this paper, we have argued that a renormalizable theory in which low-energy
scattering is characterized by a dimensionful effective coupling allows for 
only a limited range of values of boson masses, if the mass is defined as the 
distance of the complex pole from the origin. This definition is seen to be
very close to the value of the mass as derived from the peak position in the 
lineshape.
The more usual (LEP1) definition of the mass is seen to be applicable only 
over a much smaller range of parameter values.
We have studied this phenomenon using a number of approaches, culminating in 
a limit of about 1 TeV for the Higgs boson of the Minimal Standard Model.
Whereas similar limits have been obtained using arguments on the `breakdown
of perturbation theory', no such effect is either evident or necessary in our 
approach.

\paragraph{Acknowledgments} We would like to thank Pierre van Baal, Willy 
van Neerven and Jan-Willem van Holten for useful discussions.
We also acknowledge the continued interest of Ernestos Argyres and Costas 
Papadopoulos in this subject.
This research has been partly supported by the EU under contract 
number CHRX-CT-93-0319.


\end{document}